\documentclass[a4paper,11pt]{article}
\usepackage{pos}
\usepackage{bm}
\usepackage{wrapfig}

\newcommand{\ket}[1]{\left| #1 \right>} 
\newcommand{\braket}[2]{\left< #1 \vphantom{#2} \right|
 \hspace{-2pt} \left. #2 \vphantom{#1} \right>} 
\newcommand{\mbraket}[3]{\left< #1 \vphantom{#2#3} \right|
 #2 \left| #3 \vphantom{#1#2} \right>} 

\title{Quantum fluctuations of quarks and gluons in nuclei } 
\ShortTitle{Quantum fluctuations of quarks and gluons in nuclei}

\author*[a]{Michael L. Wagman}

\affiliation[a]{Fermi National Accelerator Laboratory, \\ Batavia, IL 60510, U.S.A.}

\emailAdd{mwagman@fnal.gov}

\abstract{Acceptance talk for the 2024 Kenneth G. Wilson Award for Excellence in Lattice Field Theory: \\

For key contributions to lattice QCD studies of noise reduction in nuclear systems, the structure of nuclei, and transverse-momentum dependent hadronic structure functions.}

\FullConference{The 41st International Symposium on Lattice Field Theory (LATTICE2024)\\
 28 July - 3 August 2024\\
Liverpool, UK\\}

\begin{document}
\maketitle

\section{Introduction}

This talk reviews a lattice quantum chromodynamics (LQCD) calculation of the Collins-Soper kernel governing the evolution of transverse-momentum dependent hadronic structure functions in Sec.~\ref{sec:cs}, the signal-to-noise problem facing LQCD studies of nuclear systems in Sec.~\ref{sec:nuclei}, and a new spectroscopy method based on the Lanczos algorithm and spurious eigenvalue filtering in Sec.~\ref{sec:Lanczos}. The Lanczos method mitigates analysis challenges arising from the signal-to-noise problem by providing algebraic energy estimators with asymptotically constant signal-to-noise that come with rigorous two-sided error bounds, which apply
even when excited-state effects are large.

These topics touch upon different aspects of quark and gluon quantum fluctuations in nuclei. Quantum fluctuations are responsible for both the universal correlations of lightlike Wilson lines in all hadronic and nuclear systems encoded by the Collins-Soper kernel, as well as complex phase fluctuations between Monte Carlo samples of observables in different gluon field configurations that give rise to the signal-to-noise problem.
Optimistically, one may dream that advances in our understanding of the physics of quantum fluctuations in QCD and our understanding of noise reduction in LQCD simulations will inform one another and grow together.

\section{The Collins-Soper Kernel}\label{sec:cs}

A wealth of information about the Standard Model and beyond is encoded in differential cross sections that express how interaction rates depend on the kinematics of the hadrons involved in a scattering process.
For example, while the overall rates for the Drell-Yan process $q \overline{q} \rightarrow \ell \overline{\ell} X$ and its analog $u \overline{d} \rightarrow \nu \overline{\ell} X$ depend on the $Z$- and $W$-boson masses, respectively, the event shape provides a more sensitive probe for determining $M_Z$ or $M_W$ from fits to experimental data.
In particular, the dependence of the cross section on the magnitude $q_T$ of the relative lepton momentum in the plane transverse to the collision axis has been used to provide precise determinations of $M_Z$ and $M_W$ from Tevatron~\cite{CDF:1999bpw,CDF:2003tdi,D0:2009yxq,CDF:2011ksg,CDF:2012brb,CDF:2022hxs} and LHC~\cite{CMS:2011wyd,CMS:2014cmt,CMS:2015cyj,CMS:2015zlj,CMS:2016mwa,ATLAS:2015iiu,ATLAS:2015mwq,ATLAS:2016ydt} experiments.
\begin{wrapfigure}{r}{0.43\textwidth}
\vspace{-20pt}
\begin{center}
    \includegraphics[width=0.42\textwidth]{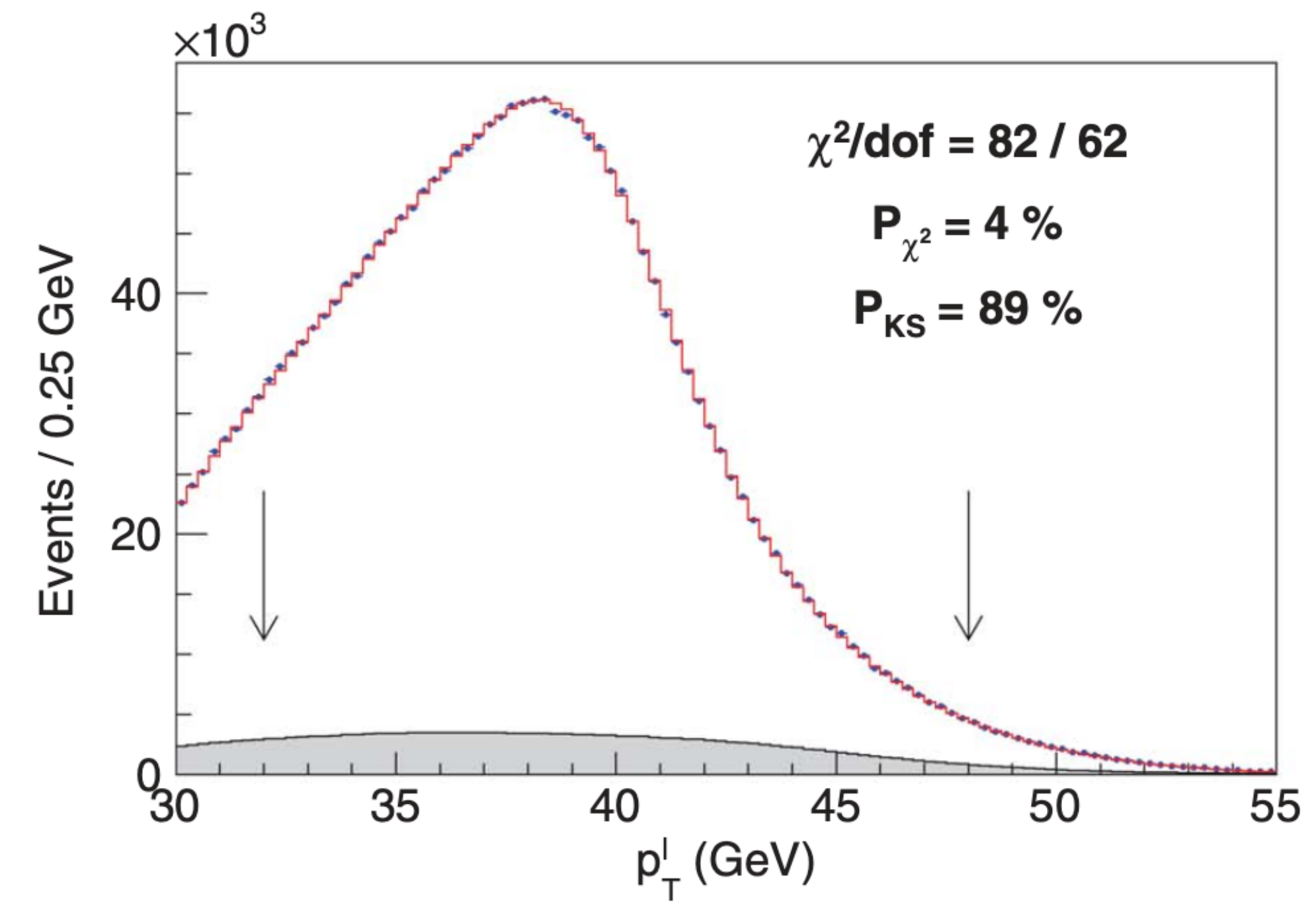}
\end{center}
\vspace{-20pt}
\caption{Tevatron event rates for muon production, $p\overline{p} \rightarrow \mu \overline{\nu} X$ / $p\overline{p} \rightarrow  \overline{\mu} \nu X$, as a function of the muon transverse momentum.  Reproduced from Ref.~\cite{CDF:2022hxs}.
\label{fig:pT}
}
\end{wrapfigure}
Fig.~\ref{fig:pT} shows an example of a transverse-momentum distribution used recently by the CDF Collaboration in a precise determination of $M_W$~\cite{CDF:2022hxs}.
Interestingly, the result shows a $7\sigma$ tension with determinations of $M_W$ based on electroweak global fits, as well as a similar tension with a more recent LHC determination by the ATLAS Collaboration~\cite{ATLAS:2024erm}.
Theoretical systematic uncertainties in $M_W$ determinations associated with higher-order perturbative QCD effects have been investigated~\cite{Isaacson:2022rts}, but quantifying theoretical uncertainties from nonperturbative QCD effects is challenging.
This motivates LQCD studies of transverse-momentum distributions, which are independently
of great interest to a variety of current and future experimental programs worldwide, including the COMPASS~\cite{Chiosso:2013ila,COMPASS:2021bws,COMPASS:2022xig,Alexeev:2022wgr,COMPASS:2023cgk,COMPASS:2023vqt,COMPASS:2023vhr} experiment at CERN, RHIC~\cite{Aschenauer:2015eha,RHICSPIN:2023zxx} at BNL, the 12~GeV program~\cite{Dudek:2012vr,CLAS:2021opg,CLAS:2021ovm,CLAS:2021lky,CLAS:2023wda,CLAS:2023gja} at the TJNAF, and the planned Electron-Ion Collider (EIC)~\cite{Boer:2010zf,Accardi:2012qut,Zheng:2018ssm,AbdulKhalek:2022hcn,Burkert:2022hjz}.

Transverse-momentum-dependent scattering rates can be described in QCD using the formalism originally proposed by Collins, Soper, and Sterman~\cite{Collins:1984kg} to factorize cross sections into hard scattering amplitudes, which can be accurately calculated in perturbative QCD for sufficiently high-energy collisions, and transverse-momentum dependent parton distribution functions (TMDPDFs).
For fixed center-of-mass energy $\sqrt{s}$, rapidity $y$, and invariant mass $Q$ of the lepton pair, this factorization takes the form
\begin{equation}\label{eq:tmd_xsec}
    \frac{d\sigma}{dq_T dQ^2 dy} = H_{j\overline{j}}(Q, \mu) \int d^2 \bm{b}_T \, e^{i\bm{q}_T \cdot \bm{b}_T}\, f_j(x_+,b_T,\mu,Q^2)\, f_{\overline{j}}(x_-,b_T,\mu,Q^2) + \ldots,
\end{equation}
where $f_j(x,b_T,\mu,\zeta)$ and $f_{\overline{j}}(x,b_T,\mu,\zeta)$ are the TMDPDFs of the partons involved in the hard scattering, $H_{j\overline{j}}(Q, \mu)$ is the hard scattering amplitude, 
and the ellipsis denotes power corrections suppressed by $q_T^2/Q^2$ and $\Lambda_{\rm QCD}^2/Q^2$.
The TMDPDFs depend on kinematic variables $x_{\pm} \equiv Q e^{\pm q} / \sqrt{s}$ and $Q^2$  
, the usual ultraviolet renormalization scale $\mu$, and an additional ``Collins-Soper'' scale $\zeta$ that must be introduced to renormalize ``rapidity divergences'' that arise in the derivation of TMD factorization~\cite{Collins:1981uk,Collins:1981va}; see for example Refs.~\cite{Collins:2014jpa,Collins:2017oxh,Ebert:2018gzl,Scimemi:2019cmh,Vladimirov:2020umg,Ebert:2022fmh} and the recent review~\cite{Boussarie:2023izj}.

The dependence of the TMDPDFs on this pair of renormalization scales is given by~\cite{Collins:1981uk,Collins:1981va}
\begin{equation}
    f_j(x,b_T,\mu,\zeta) = f_j(x,b_T,\mu_0,\zeta_0) \exp\left[ \int_{\mu_0}^\mu \frac{d \mu' }{\mu'} \gamma_\mu^j(\mu',\zeta_0) \right] \exp\left[ \frac{1}{2} \gamma_\zeta^j(\mu,b_T) \ln \frac{\zeta}{\zeta_0} \right].
\end{equation}
Knowledge of both the ultraviolet anomalous dimension $\gamma_\mu^j(\mu',\zeta_0)$ as well as the ``Collins-Soper kernel,'' also often called the ``rapidity anomalous dimension,'' $\gamma_\zeta^j(\mu,b_T)$ is required in order to relate the TMDPDFs at a fixed reference scale---which can either be related perturbatively to (collinear) PDFs or treated as nonperturbative inputs to be fit to data---to those at other scales.
Commutation of $\mu$ and $\zeta$ derivatives links the ultraviolet and rapidity anomalous dimensions as
\begin{equation}\label{eq:CS_dmu}
    \mu \frac{d}{d\mu} 2 \zeta \frac{d}{d\zeta} f_j(x,b_T,\mu,\zeta)  =  \mu \frac{d}{d\mu} \gamma_\zeta^j(\mu, b_T) = 2 \zeta \frac{d}{d\zeta} \gamma_\mu^j(\mu,\zeta) \equiv -2 \Gamma_{\rm cusp}^j(\alpha_s(\mu)),
\end{equation}
where $\Gamma_{\rm cusp}$ is the ``cusp anomalous dimension''  that governs the renormalization of Wilson loops with corners~\cite{Polyakov:1980}. 
It is a universal quantity relevant to e.g.~back-to-back jets that are described by Wilson lines at nearly opposite angles, as well as matrix elements of nonlocal quark operators involving staple-shaped Wilson lines or other configurations with corners, and has now been computed in perturbative QCD at two, three, and four loops~\cite{Korchemsky:1987wg,Moch:2004pa,Henn:2019swt}.
A consequence of Eq.~\eqref{eq:CS_dmu} is that the ultraviolet anomalous dimension can be expressed as
\begin{equation}
    \gamma_\mu^i(\mu, \zeta) = \ln(\mu^2 / \zeta) \Gamma_{\rm cusp}^j(\alpha_s(\mu)) + \gamma_\mu^i(\alpha_s(\mu)).
\end{equation}
The key feature of this expansion is the $\alpha_s(\mu)$ is always evaluated at the ultraviolet renormalization scale $\mu$, which should be of order $\mu \sim Q$ to avoid large logarithms.
Therefore, both the cusp and non-cusp ultraviolet anomalous dimension can be accurately calculated using perturbative QCD as long as $Q \gg \Lambda_{\rm QCD}$;
the non-cusp part has been computed at two loops~\cite{Becher:2006mr}.
On the other hand, the Collins-Soper (CS) kernel can be expressed as
\begin{equation}
    \gamma_\zeta^i(\mu, \zeta) = -2 \int_{1/b_T}^\mu \frac{d\mu'}{\mu'} \Gamma_{\rm cusp}^j(\alpha_s(\mu)) + \gamma_\zeta^i(\alpha_s(1/b_T)).
\end{equation}
Both the cusp and non-cusp parts of the CS kernel involve $\alpha_s$ evaluated at scales $\geq 1/b_T$ and can be accurately calculated in perturbative QCD only if $1/b_T \ll \Lambda_{\rm QCD}$.
The non-cusp part has been computed at three and four loops~\cite{Li:2016ctv,Duhr:2022yyp}, the same accuracy as the cusp part.

The Fourier transform in Eq.~\eqref{eq:tmd_xsec} leads to integral contributions from $b_T \sim 1/q_T$ dominating the cross-section integral, Eq.~\eqref{eq:tmd_xsec}.
This means that even for $Q \gg \Lambda_{\rm QCD}$, nonperturbative QCD effects are important in the CS kernel whenever $q_T / Q$ is sufficiently small.
Shapes of transverse-momentum distributions, in particular the heights and widths of the ``Sudakov peaks'' where event rates are largest (see Fig.~\ref{fig:pT}), vary with $Q^2$ in a way that is governed by the CS kernel.
Nonperturbative effects in the CS kernel have been studied phenomenologically through global fits to experimental data. 
These fits are complicated by the fact that experimental data depend on transverse momenta $q_T$, while the CS kernel is intrinsically a function of the impact parameter $b_T$, and by the fact that the fixed-scale TMDPDFs $f_j(x,b_T,\mu_0,\zeta_0)$ are nonperturbative inputs that must be fit simultaneously.
Despite these challenges, global fits to semi-inclusive deep inelastic scattering (SIDIS) and Drell-Yan data have been performed by several groups~\cite{Landry:2002ix,Scimemi:2019cmh,Bacchetta:2019sam,Bacchetta:2022awv,Moos:2023yfa,Isaacson:2023iui}.
Results for the CS kernel from these global fits are in good agreement for $b_T \lesssim 0.2$ fm; however, discrepancies between fits to different CS kernel parameterizations appear for larger $b_T$ that indicate that the nonperturbative region is not adequately constrained by current experimental data.

Another important feature of both TMDPDF anomalous dimensions is that they are independent of the hadron state involved in the scattering process~\cite{Collins:1981uk,Collins:1981va,Collins:1984kg}.
This can be seen from the matrix element definition of TMDPDFs~\cite{Collins:1984kg,Echevarria:2011epo,Chiu:2012ir,Ebert:2018gzl},
\begin{equation}
    f_j(x,b_T,\mu,\zeta) = \lim_{\epsilon,\tau \rightarrow 0} Z(\mu, \zeta, \epsilon) B_j(x,\bm{b}_T,\epsilon,\tau,\zeta) \Delta_j(b_T, \epsilon, \tau),
\end{equation}
where the ``beam function'' $B_j(x,\bm{b}_T,\epsilon,\tau,\zeta)$ is defined as a matrix element of a nonlocal quark bilinear operator involving a staple-shaped Wilson line, $\epsilon$ is the dimensional regularization cutoff, $\tau$ is the rapidity regulator, 
and $\Delta_j(b_T, \epsilon, \tau)$ is a ``soft factor'' defined from a vacuum matrix element of a Wilson-loop operator involving lightlike Wilson lines with both past and future orientations.
Cancellation of $1/\tau$ divergences is ensured by the factor that $B_j(x,\bm{b}_T,\epsilon,\tau,\zeta)$ only depends on the combination $1/\tau - \ln \sqrt{\zeta}$ for small $\tau$~\cite{Ebert:2018gzl}. 
This fact relates $\frac{d}{d\zeta} f_j$ to $\frac{d}{d\tau} \Delta_j$, which only involves a QCD vacuum matrix element.
The CS kernel can thus be viewed as describing how the correlations of nearly lightlike gluonic fluctuations of the QCD vacuum depend on their rapidity, 
and must be independent of the hadron state used to define $f_j$.

Due to the hadron state independence of the CS kernel, the same nonperturbative QCD effects govern the shapes of transverse-momentum distributions for collisions involving nucleons and nuclei.
These effects are independent of the atomic number or state of the nucleus and apply to both hadron-hadron and lepton-hadron collisions.
For example, current and future accelerator neutrino experiments such as NOvA, T2K, MicroBooNE, ICARUS, SBND, DUNE, and Hyper-Kamiokande, require the energy dependence of neutrino-nucleus cross sections as inputs in order to determine neutrino oscillation parameters.
Events involving pion production are common in current experiments, and will be even more so at DUNE, and the transverse-momentum dependence of pion production rates in low-energy analogs of SIDIS provide an important input needed to analyze these events.
The CS kernel governs the shapes of Sudakov peaks for transverse-momentum distributions in e.g.~charged-current-single-pion-production (CC1$\pi$) events.
This links the cross-section needs of neutrino experiments to quantities that can be accessed through hadron-hadron or lepton-hadron colliders like the LHC and future EIC.
Accurate nonperturbative determination of the CS kernel from QCD therefore provide a universal ingredient for analyzing nuclear scattering processes in a wide range of experiments.
Such predictions allow global fits of nonperturbative hadron and nuclear structure functions to focus on fixed-scale TMDPDFs without needing to also fit nonperturbative evolution effects, a situation more analogous to global fits of collinear PDFs whose perturbative evolution is known perturbatively.

Direct calculation of the lightlike Wilson line operators involved in the definitions of TMDPDFs and the CS kernel is obstructed by the fact that LQCD calculations are necessarily performed in Euclidean spacetime.
This obstacle can be circumvented using the large momentum effective theory (LaMET) framework~\cite{Ji:2013dva} as first proposed in the context of TMDPDFs by Xiangdong Ji~\cite{Ji:2014hxa}.
The main idea, as reviewed in Ref.~\cite{Ji:2020ect}, is that matrix elements of spacelike-separated quark bilinears in highly boosted hadron states approach those of lightlike-separated operators up to logarithmic corrections that can be accurately calculated in perturbation theory.
For the CS kernel in particular, additional obstacles arise from the pair of orthogonal lightlike directions appearing in the soft function~\cite{Ebert:2019okf}.
It has recently been shown how to explicitly determine the soft factor from highly boosted form factors with asymptotically large momentum transfer~\cite{Ji:2019sxk}, but for the purposes of determining the CS kernel in particular it is convenient to form ratios of beam functions for which the soft factor cancels between numerator and denominator.
A formula for obtaining the quark CS kernel from ratios of Euclidean matrix elements convolved with perturbative LaMET matching factors was explicitly given by Ebert, Stewart, and Zhao~\cite{Ebert:2018gzl}:
\begin{equation}\label{eq:ESZ}
    \gamma_\zeta^q(\mu, b_T) = \frac{1}{\ln(P^z_1/P^z_2)} \ln \left[ \frac{C(\mu, xP^z_2) \int d b^z e^{i x b^z P^z_1} B_q^{\overline{\text{MS}}}(\bm{b}_T, b^z, \ell, P^z_1, \mu)}{C(\mu, xP^z_1) \int d b^z e^{i x b^z P^z_2} B_q^{\overline{\text{MS}}}(\bm{b}_T,  b^z, \ell,  P^z_2, \mu) } \right],
\end{equation}
where the Euclidean quasi beam functions are defined by
\begin{equation}\label{eq:beam}
    B_q^{\overline{\text{MS}}}(\bm{b}_T, b^z, \ell, P^z_1, \mu) = \sum_{\Gamma'} Z_{\Gamma \Gamma'}^{\overline{\text{MS}}}(\mu, a) \left\langle h(P^z) \vphantom{\frac{1}{2}} \right| \overline{q} \left(\frac{b^z}{2}\hat{\bm{e}}_z + \bm{b}_T \right) \Gamma' W_\sqsupset(b^z,\bm{b}_T,\ell) q\left(-\frac{b^z}{2}\hat{\bm{e}}_z\right) \left|  \vphantom{\frac{1}{2}} h(P^z) \right\rangle
\end{equation}
for a particular hadron state $\left|h(P^z)\right\rangle$ with momentum $\bm{P} = P^z \hat{\bm{e}}_z$; the Dirac matrix $\Gamma$ can be chosen to be any linear combination of $\gamma_z$ and $\gamma_t$;
and the staple-shaped Wilson line is defined as
\begin{equation}
    W_\sqsupset(b^z,\bm{b}_T,\ell) = W\left[ \bm{b}_T + \left(\frac{b^z}{2}\right) \hat{\bm{e}}_z,\, \left( \frac{\ell}{2} \right) \hat{\bm{e}}_z \right]
    W\left[ \left( \frac{\ell}{2} \right) \hat{\bm{e}}_z + \bm{b}_T,\, \left( \frac{\ell}{2} \right) \hat{\bm{e}}_z \right] 
    W\left[ \left( \frac{\ell}{2} \right) \hat{\bm{e}}_z,\, \left(-\frac{b^z}{2}\right) \hat{\bm{e}}_z \right],
\end{equation}
where $W[A,B]$ is a Wilson line from $A$ to $B$ and 
$Z_{\Gamma \Gamma'}^{\overline{\text{MS}}}(\mu, a)$ are renormalization factors that depend on the lattice spacing $a$.
The renormalization of nonlocal quark bilinear operators is more complicated than that of local operators and has been studied extensively in recent 
years~\cite{Ji:2017oey,Green:2017xeu,Constantinou:2017sej,Constantinou:2019vyb,Shanahan:2019zcq,Ebert:2019tvc,Green:2020xco,Zhang:2020rsx,LatticePartonLPC:2021gpi,Zhang:2022xuw,Alexandrou:2023ucc,Zhang:2024omt}.
I will not attempt to review this progress here and simply state one important result: interpreting Wilson lines as propagators 
of auxiliary static quark fields provides an interpretation of 
nonlocal quark bilinears as products of (possibly many) local operators~\cite{Ji:2017oey,Green:2017xeu,Green:2020xco}.
For example, this staple-shaped Wilson line 
can be expressed as a product of two static-light transition operators $\overline{Q}_z q$ and $\overline{q} Q_z$, as well as two static-static transition operators $\overline{Q}_\perp Q_z$ and $\overline{Q}_z Q_\perp$, involving auxiliary fields $Q_z$ and $Q_\perp$ propagating in the $z$ and transverse directions, respectively~\cite{Green:2020xco}.
The renormalization factors for each of these local currents can be then computed in a variation of commonly used nonperturbative momentum subtraction schemes.
This ``RI-xMOM'' scheme~\cite{Ji:2017oey,Green:2017xeu,Green:2020xco} provides a conceptually simple and robust way to approach renormalization of non-local operators.

\begin{figure}[t]
                \centering
                \includegraphics[width=0.42\textwidth]{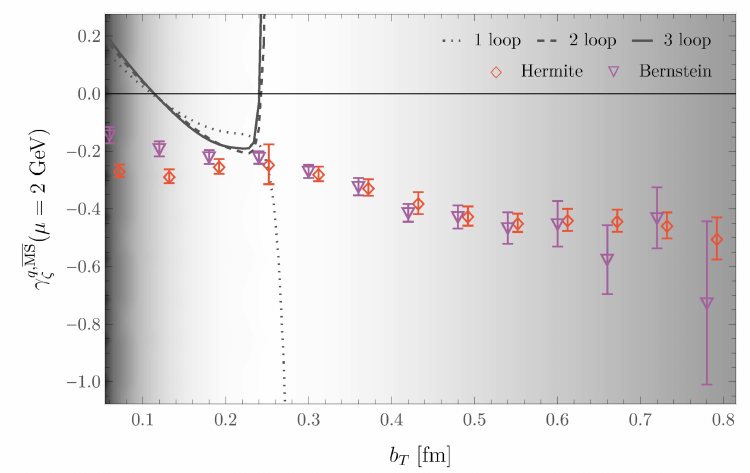}
                \includegraphics[width=0.42\textwidth]{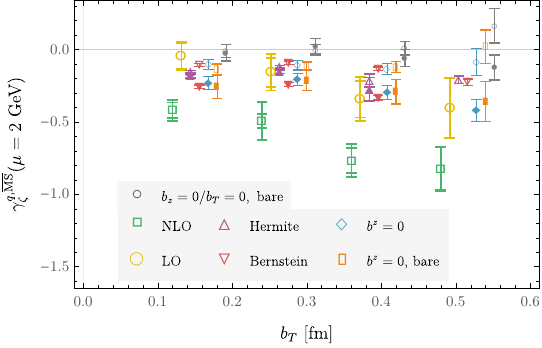}
                 \includegraphics[width=0.42\textwidth]{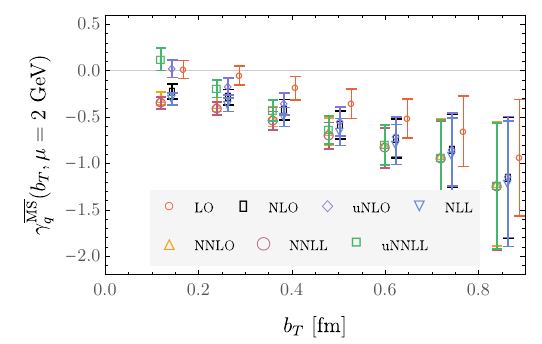}
                \includegraphics[width=0.42\textwidth]{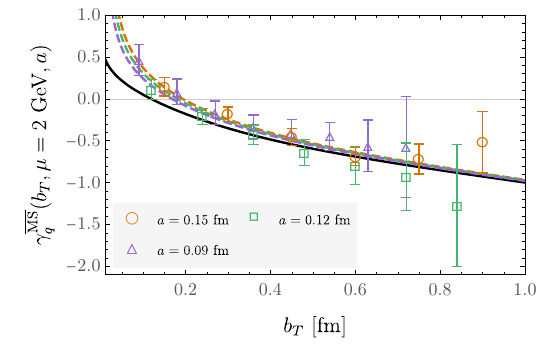}
            \caption{\label{fig:lattcomparison}
                Evolution of LQCD determinations of the CS kernel from 2020-2024. (Top-left) exploratory quenched QCD results~\cite{Shanahan:2020zxr}, (top-right) partially quenched results demonstrating the inadequacy of previously used approximations~\cite{Shanahan:2021tst}, (bottom-left) physical-quark-mass results showing the importance of power-suppressed matching effects at small $b_T$~\cite{Avkhadiev:2023poz}, and (bottom-right) results at multiple lattice spacings~\cite{Avkhadiev:2024mgd}.
                }
\end{figure}

Calculation of the CS kernel using LQCD is greatly facilitated by its hadron state independence.
Although nucleons and nuclei are the most experimentally relevant systems, any state can be used to study the universal QCD vacuum fluctuations encoded in the CS kernel.
Shortly after Eq.~\eqref{eq:ESZ} was presented, Phiala Shanahan, Yong Zhao, and I set out to compute the CS kernel using the simplest hadron state in LQCD: the pion.
We began in quenched QCD, where state independence implies that the CS kernel can be determined using valence quark probes with any mass.
Using $m_\pi \sim 1$ GeV allowed precise signals for the CS kernel at $b_T$ in the nonperturbative region to be obtained using very modest computational resources. 
Our final results from this study~\cite{Shanahan:2020zxr}
are shown in Fig.~\ref{fig:lattcomparison}; in comparison to more recent results it is clear that discretization effects are significant at small $b_T$, but the large-$b_T$ behavior is surprisingly similar to state-of-the-art unquenched results.

Several challenges for determining the CS kernel more precisely were identified in Ref.~\cite{Shanahan:2020zxr}.
Some were related to renormalization issues discussed above.
Another is the need to fit a large number of bare matrix elements for different staples, boosts, and Dirac structures---35,000 matrix elements for this exploratory calculation---which required the introduction of robust automated fitting methods.
The other important challenge is accurately performing the Fourier transform in Eq.~\eqref{eq:ESZ}. 
The matrix elements entering the quasi beam function decay exponentially for large $b^z P^z$; however, broadening of the beam function with increasing $b_T$ leads to requirements that larger $b^z P^z$ must be achieved for large-$b_T$ TMDPDF Fourier transforms than for analogous quasi PDF calculations.
Large enough values of $b^z P^z$ were not possible to achieve for the lattice volume used in Ref.~\cite{Shanahan:2020zxr}, and the Fourier transform was instead performed using a fitted model of the beam function that is only consistent with tree-level light-cone matching in Eq.~\eqref{eq:ESZ}.
Other early LQCD calculations of the CS kernel by the LPC Collaboration~\cite{LatticeParton:2020uhz}, ETMC/PKU~\cite{Li:2021wvl}, and the Regensburg group~\cite{Schlemmer:2021aij} also employed approximations only consistent with tree-level matching.

We studied these and other approximations in detail in Ref.~\cite{Shanahan:2021tst}, where higher-statistics studies on a larger lattice volume enabled better-controlled Fourier transforms that could be combined with one-loop matching.
As shown in Fig.~\ref{fig:lattcomparison}, one-loop matching effects~\cite{Ebert:2018gzl,Ebert:2019okf} are significant and tree-level  approximations are inadequate for a high-precision determination. 
Further progress was made by the LPC Collaboration in Ref.~\cite{LPC:2022ibr}, where it was realized that the CS kernel can be extracted from quasi TMD wavefunctions related to pion-to-vacuum transition matrix elements of the same staple-shaped operators discussed above.
Such TMD wavefunctions can be determined from two-point correlation functions that are less computationally expensive than the three-point functions required to compute quasi beam functions.
Using high-statistics calculations of quasi TMD wavefunctions, Ref.~\cite{LPC:2022ibr} obtained results at
 larger $b^z P^z$ with much smaller Fourier transform truncation effects.
TMD wavefunctions were adopted in subsequent calculations~\cite{Shu:2023cot,LatticePartonLPC:2023pdv,Avkhadiev:2023poz,Avkhadiev:2024mgd}.

The first LQCD calculation of the CS kernel employing approximately physical quark masses and performing a continuum extrapolation using results at multiple lattice spacings was reported in Refs.~\cite{Avkhadiev:2023poz,Avkhadiev:2024mgd}.
Large enough boosts and staple extents were used that Fourier transform truncation effects could be explicitly computed and verified to be negligible.
Perturbative matching was performed at two-loop level based on results from Refs.~\cite{Ji:2021uvr,Deng:2022gzi,delRio:2023pse,Ji:2023pba}, and resummation of momentum logarithms is performed using results of Refs.~\cite{Ji:2019ewn,Ebert:2022fmh}.
The dependence on perturbative order is observed to be mild at large $b_T$; however, significant differences between different orders are visible in Fig.~\ref{fig:lattcomparison} at small $b_T$.
To partially account for power-suppressed $1/(P^z b_T)$ effects that are formally negligible but numerically significant at computationally accessible momenta and lattice spacings, an unexpanded matching scheme was proposed in  Ref.~\cite{Avkhadiev:2023poz} that should reduce higher-order effects at finite $P^z$ and small $b_T$.
Differences between tree-level, (resummed) one-loop, 
and (resummed) two-loop results are observed to be smaller in this scheme, and the 
\begin{wrapfigure}{r}{0.47\textwidth}
\vspace{-20pt}
\begin{center}
    \includegraphics[width=0.47\textwidth]{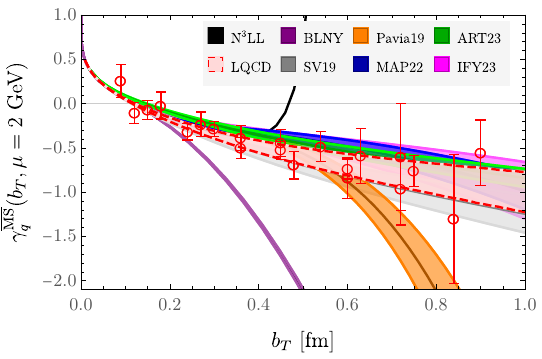}
\end{center}
\vspace{-20pt}
             \caption{\label{fig:phenocomparison}
                Lattice QCD parameterization of the CS kernel, phenomenological parameterizations~\cite{Landry:2002ix,Scimemi:2019cmh,Bacchetta:2019sam,Bacchetta:2022awv,Moos:2023yfa,Isaacson:2023iui} of experimental data (BLNY, SV19, Pavia19, MAP22, ART23, IFY23), and perturbative results from Refs.~\cite{Collins:2014jpa,Li:2016ctv,Vladimirov:2016dll} ($\mathrm{N}^3\mathrm{LL}$). Reproduced from Ref.~\cite{Avkhadiev:2024mgd}.
                 }\vspace{-10pt}
\end{wrapfigure}
unexpanded next-to-next-to-leading-log (uNNLL) scheme is used for final results in Refs.~\cite{Avkhadiev:2023poz,Avkhadiev:2024mgd}.

Discretization effects were observed to be significant at small $b_T$ in Ref.~\cite{Avkhadiev:2024mgd}, which used three lattice spacings $a \in \{0.09,0.12,0.15\}$ fm in a mixed-action setup with clover valence fermions and gauge-fields generated by the MILC collaboration~\cite{MILC:2012znn}, see Fig.~\ref{fig:lattcomparison}.
These discretization effects were fit along with a parameterization of the nonperturbative part of the CS kernel,
\begin{equation}\label{eq:CSkernelparam}
\begin{aligned}
    \gamma_\zeta^q(b_T, \mu, a) =  -2\mathcal{D}(b^*,\mu)  -2 c_0 b_T b^* + k_1 \frac{a}{b_T},
    \end{aligned}
\end{equation}
where $b^* = b_T/ \sqrt{1+ b_T^2/B^2_\text{NP}}$ with $B_{\text{NP}} = 2$ GeV and $\mathcal{D}(b^*,\mu)$ is the resummed four-loop
perturbative expression for the CS kernel presented explicitly in Ref.~\cite{Avkhadiev:2024mgd}.
This parameterization 
provides a good description of LQCD results with $b_T \in [0.09, 0.9]$ fm and $a \in [0.09,0.15]$ fm with best-fit parameters $c_0 = 0.032(12)$ and $k_1 = 0.22(8)$ achieving $\chi^2 / \text{dof} = 0.39$.
Additional terms, e.g.~proportional to $\ln(b_*)$ and $a^2 / b_T^2$, do not significantly improve the fit quality.
Other parameterizations of the nonperturbative part, including the BLNY parameterization~\cite{Landry:1999an} and the hadron-structure-oriented (HSO)  parameterization introduced in Ref.~\cite{Aslan:2024nqg}, provide equally accurate descriptions for $b_T \lesssim 1$ fm after fitting to LQCD results.
The continuum-extrapolated CS kernel parameterization corresponding to Eq.~\eqref{eq:CSkernelparam} with $c_0 = 0.032(12)$ and $k_1=0$  is the final result of these calculations and provides the first nonperturbative QCD determination of the CS kernel with fully quantified uncertainties.
As seen in Fig.~\ref{fig:phenocomparison}, LQCD results are consistent with the most recent global fit results and are precise enough to exclude some older fits.
LQCD results are more precise than state-of-the-art global fits across the range $b_T \lesssim 1$ fm constrained by data.
In future work, they will be incorporated into global fits and cross-section predictions for Drell-Yan, SIDIS, and other processes.

\section{From QCD to Nuclei}\label{sec:nuclei}

Most features of scattering cross sections are not as universal as the evolution effects governed by the Collins-Soper kernel and must be computed for the particular hadrons and nuclei involved in experiments.
In particular, the study of nuclei has been a long-standing goal for LQCD calculations.
A major challenge for studying nuclei is the ``signal-to-noise problem'': the signal-to-noise (SNR) ratios of nucleon and nuclear correlation functions (correlators) decrease exponentially with increasing Euclidean time.
The physical reason for this SNR degradation was elucidated by Parisi~\cite{Parisi:1983ae} and Lepage~\cite{Lepage:1989hd} by studying correlator variances as physical correlation functions themselves.
For a nucleon correlation function proportional to $e^{-M_N t}$ at large Euclidean times $t$, the variance includes not only contributions from nucleon-antinucleon states $\propto e^{-2M_Nt}$ but also contributions from three-pion states $\propto e^{-3m_\pi t}$ with the same quantum numbers\footnote{Because quark-field integrals are performed analytically, the variance correlator should formally be defined in a partially quenched theory where valence quark and antiquark numbers are separately conserved. This excludes single-pion states with the same physical quantum numbers from the nucleon variance correlator.} that decay exponentially slower.
This implies that the SNR for nucleon correlators, defined as the average correlator divided by the square root of its variance, decays exponentially with Euclidean time with a rate of $M_N - 3m_\pi/2$.
This exponential SNR degradation becomes increasingly severe for multi-nucleon systems, where systems of $A$ nucleons have variance correlators with $3A$-pion ground states.
Since nuclear binding energies per nucleon, ranging between 1-8 MeV, are less than a percent of the nucleon mass, the SNR expectations for $A$ non-interacting nucleons provide a good approximation to the asymptotic scaling expected for nuclear correlators, whose SNR exponentially decays at a rate of $A(M_N - 3m_\pi/2)$~\cite{Beane:2009gs}.

Quark operators are charged under the $U(1)_B$ baryon-number symmetry of QCD and quark propagators have complex phases that shift under $U(1)_B$ transformations, see e.g.~Ref.~\cite{Ce:2016idq}.
These phases depend on the gauge-field background and encode physical information analogous to Aharonov-Bohm phases.
This means that the phases of baryon correlators, as well as the signs of meson correlators besides the pion, have gauge-field-dependent phase fluctuations that lead to ``sign problems'' for correlators~\cite{Wagman:2016bam,Wagman:2017gqi}.
The average phase factor $\left< C_N(t) / |C_N(t)| \right>$ for a nucleon correlator $C_N(t) \equiv \left< N(t) \overline{N}(0) \right>$ is empirically observed to be proportional to $e^{-(M_N - 3m_\pi/2)t}$ for large Euclidean times, demonstrating that this sign problem can also be viewed as responsible for the nucleon SNR problem~\cite{Wagman:2016bam}.
Similar features appear for meson and nuclear correlators~\cite{Wagman:2017gqi,Davoudi:2020ngi}.

This observation has led to the application of new methods for improving correlator SNR based on applying methods to correlators~\cite{Detmold:2018eqd,Detmold:2020ncp,Detmold:2021ulb,Lin:2023svo} that had been previously applied to sign problems associated with finite-density systems~\cite{Ejiri:2007ga,Nakagawa:2011eu,Cristoforetti:2012su,Cristoforetti:2013wha,Aarts:2013fpa,Alexandru:2015sua,Alexandru:2017czx,Mori:2017nwj,Alexandru:2020wrj}.
Techniques based on path integral contour deformations can provide orders-of-magnitude improvements to correlator SNR in two-dimensional gauge theories~\cite{Detmold:2020ncp,Detmold:2021ulb} but achieving large SNR gains in systems with three or more dimensions has proven more challenging~\cite{Lin:2023svo}.
Understanding of the statistical distributions of correlator phases has also improved through the recognition that large-time distributions of (real and imaginary parts of) nucleon and nuclear correlators are qualitatively described by complex log-normal distributions~\cite{Wagman:2016bam,Wagman:2017jva,Davoudi:2020ngi} in a similar way that the early-time distributions of correlator real parts is ubiquitously log-normal~\cite{Guagnelli:1990jb,Endres:2011jm,DeGrand:2012ik}.
Recent calculations of the exact distributions of correlator fluctuations in scalar field theories~\cite{Yunus:2022wuc,Yunus:2023dka} demonstrate that precise knowledge of this distribution can assist in the construction of more precise estimators than the sample mean, which 
\begin{wrapfigure}{r}{0.43\textwidth}
\vspace{-20pt}
\begin{center}
    \includegraphics[width=0.42\textwidth]{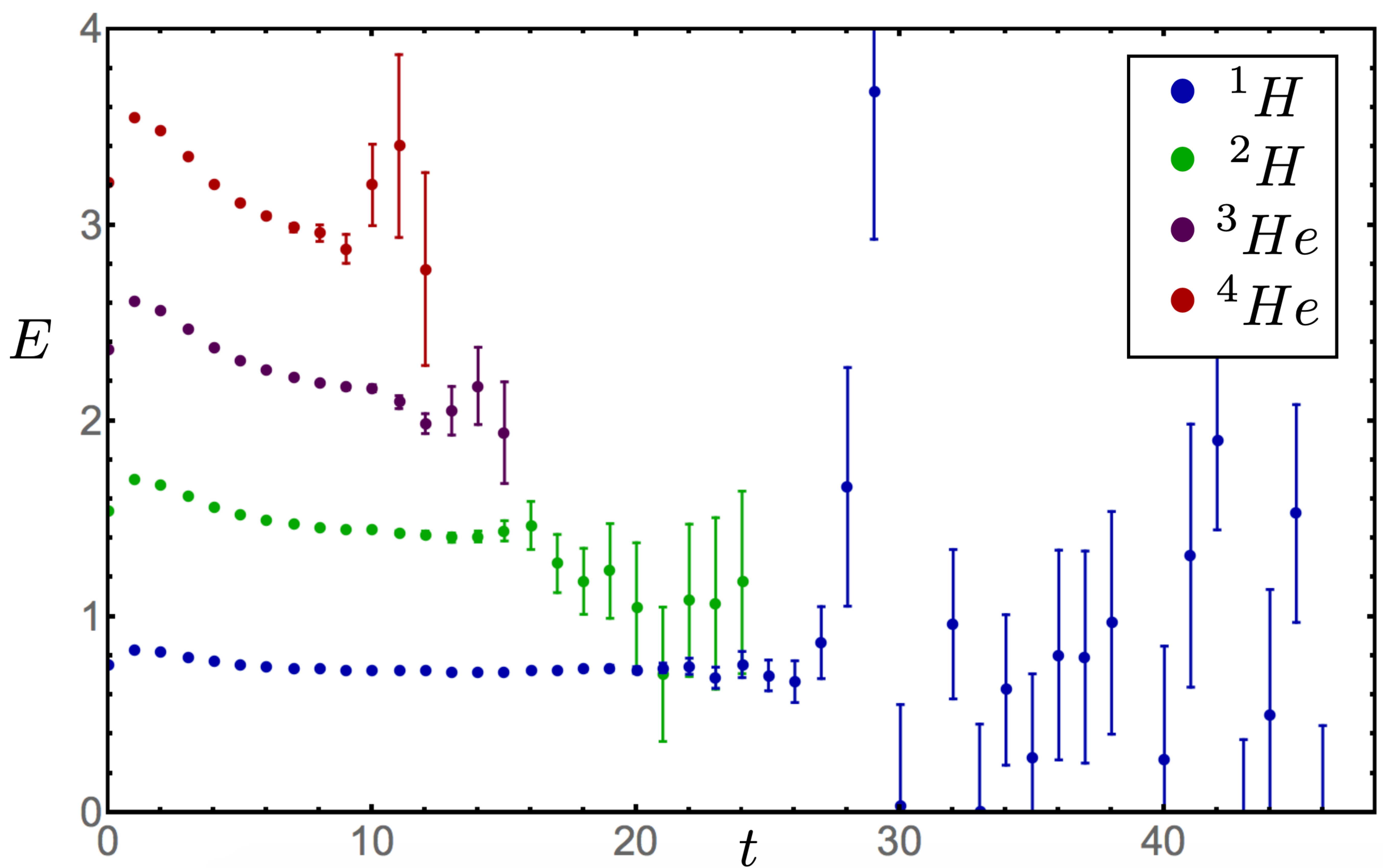}
\end{center}
\vspace{-20pt}
\caption{Effective masses for light nuclei with $A \in \{1,2,3,4\}$ from LQCD calculations with $m_\pi \sim 450$ MeV;  data from Refs.~\cite{Orginos:2015aya,NPLQCD:2020lxg}.
\label{fig:eff_mass_vs_A}
} 
\end{wrapfigure}
might also exist for correlators in LQCD.

The same rate of SNR degradation applying to nuclear correlators is inherited by standard ground-state energy estimators such as the effective mass
\begin{equation}\label{eq:eff_mass}
    E(t) = -\frac{1}{a}\left[ \ln C(t) - \ln C(t-a) \right] = E_0 + O(e^{-\delta t}),
\end{equation}
which asymptotically approaches the ground-state energy plus exponentially-suppressed corrections.
This increasing rate of SNR degradation with $A$ is exhibited in calculations of nuclear 
correlators
with $A\in \{1,2,3,4\}$ computed
by the NPLQCD Collaboration displayed in Fig.~\ref{fig:eff_mass_vs_A}.
This SNR degradation rate is especially problematic for nuclei, where the excitation gap $\delta$ controlling the size of excited-state effects is on the few MeV scale of binding energies rather than the hundreds of MeV scales associated with hadron resonance excitation energies or energy gaps to states containing extra pions.
This means that imaginary times of hundreds of fm would be required to achieve $\delta t \gg 1$ and thus meaningfully suppress excited-state effects in nuclear correlators; however SNR degradation limits the range of times where reliable signals can be extracted in calculations feasible with current resources to imaginary times of a few fm.
Interpreting effective energies, or multi-state fits to correlators with $t \ll$ 100 fm, as reliable determinations of nuclear ground-state energies therefore requires physical arguments that ground-state overlaps in a particular correlator are larger than all excited-state overlaps.

Exploratory LQCD calculations of nuclei have mostly used unphysically heavy quark masses to reduce the severity of the SNR problem. The empirical formula $M_N \sim 800$ MeV $+ m_\pi$~\cite{Walker-Loud:2014iea} shows that SNR degradation scales as $M_N - \frac{3m_\pi}{2} \sim 800 \text{ MeV} - \frac{m_\pi}{2}$ and is half as severe for $m_\pi \sim 800$ MeV as for the chiral limit.
However, a downside of beginning with calculations using unphysically heavy quark masses is that the physical spectrum is not known \emph{a priori} and is a genuine prediction.

\begin{figure}[t]

                \centering
                \includegraphics[width=0.85\textwidth]{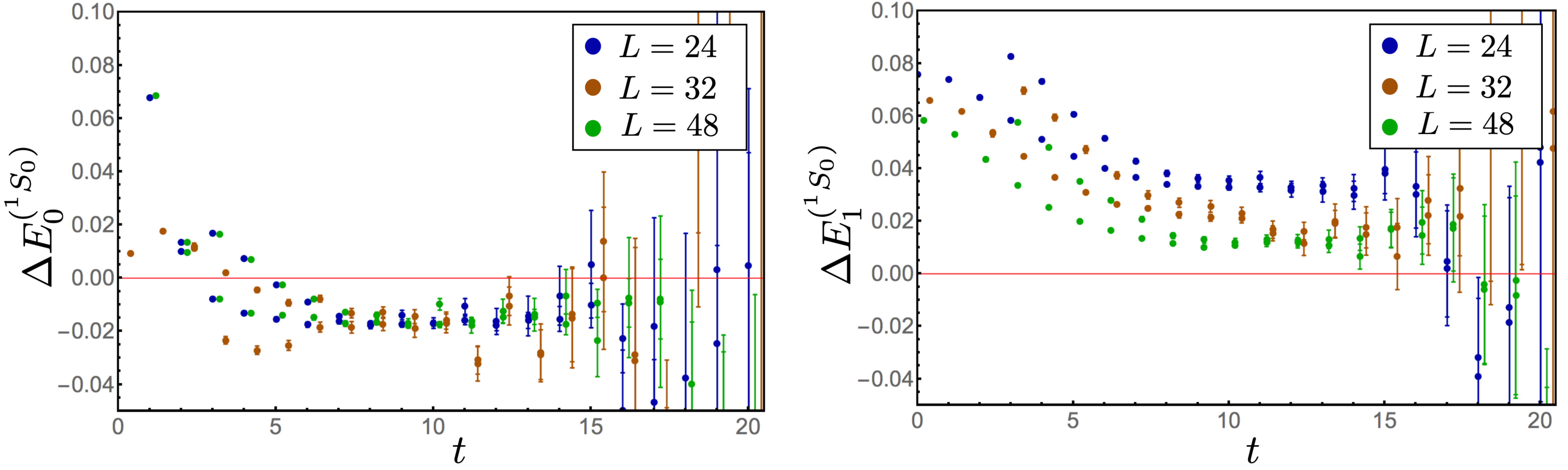}
            \caption{\label{fig:eff_mass_NN}
               Two-neutron effective FV energy shifts with $m_\pi \sim 800$ MeV, defined as differences between dineutron effective masses and twice the neutron effective mass. Data are from Ref.~\cite{Wagman:2017tmp}.
                }\vspace{-10pt}
\end{figure}

The first generation of LQCD calculations of nuclear correlators~\cite{Beane:2006mx,Beane:2009py,NPLQCD:2011naw,NPLQCD:2012mex,NPLQCD:2013bqy,Yamazaki:2012hi,Yamazaki:2015asa,Orginos:2015aya,Berkowitz:2015eaa} was made possible by efficient algorithms~\cite{Doi:2012xd,Detmold:2012eu} for computing correlators, in particular where all quark fields were placed at the same point at the source.
This source does not resemble the eigenstates of non-interacting nucleons in a finite-volume (FV), which in the rest frame are products of single nucleon fields of the form $N(\bm{p}) N(-\bm{p})$ with FV momenta $\bm{p} = (2\pi/L) \bm{n}$ quantized in terms of integers $\bm{n} \in \mathbb{Z}^3$ and energies $2 \sqrt{M_N^2 + (2\pi \bm{n}/L)^2}$.
By constructing asymmetric correlators where the sink fields were projected into products of plane-wave baryon interpolators analogous to  $N(\bm{p}) N(-\bm{p})$, it was hoped that the overlap to other FV eigenstates resembling two nucleons with relative momentum $\bm{k} \neq \bm{p}$ would be suppressed by factors proportional to the inverse spatial volume that would effectively suppress excited-state effects even for computationally achievable $t \ll \delta^{-1}$.
The degree of overlap-factor suppression achieved is difficult to estimate \emph{a priori}, and therefore early calculations performed a number of \emph{a posteriori} checks on ground-state saturation.
One necessary condition is that different interpolating operators reach the same effective-mass ``plateau,'' which is observed when comparing asymmetric correlators with Gaussian-smeared vs point-like sources, see Fig.~\ref{fig:eff_mass_NN}.
Another necessary condition is consistency between results using different physical volumes, for which the FV spectra of excited states will be very different.
This consistency was also observed in early calculations, where the appearance of an approximately volume-independent plateau below the two-nucleon threshold for $N(\bm{0})N(\bm{0})$ sinks is consistent with physical expectations for a two-nucleon bound state, and the appearance of clear volume dependence in a plateau for $N(\bm{1}\  2\pi /L)N(\bm{-1}\  2\pi/L)$ sinks is consistent with physical expectations for an unbound scattering state~\cite{NPLQCD:2013bqy,Orginos:2015aya}. 
Further checks were performed by investigating the scattering amplitudes~\cite{Iritani:2017rlk,Wagman:2017tmp,NPLQCD:2020lxg} associated with the FV spectrum through generalizations of L{\"u}scher's FV quantization condition~\cite{Luscher:1986pf,Luu:2011ep,Briceno:2013bda}.
Pionless effective field theory (EFT) was tuned to match early LQCD results for two- and three-nucleon energy spectra and then postdicted four-nucleon binding energies that were consistent with direct LQCD calculations~\cite{Barnea:2013uqa}; providing another validation of both methods.
Nuclear matrix elements of vector~\cite{Beane:2014ora,Beane:2015yha,Chang:2015qxa,Detmold:2015daa,Parreno:2016fwu}, axial~\cite{Savage:2016kon,Shanahan:2017bgi,Tiburzi:2017iux,Parreno:2021ovq,Davoudi:2024ukx}, scalar~\cite{Chang:2017eiq}, and tensor~\cite{Chang:2017eiq} currents, as well as moments of nuclear PDFs~\cite{Winter:2017bfs,Detmold:2020snb}, were subsequently computed using this asymmetric correlator setup and provide results in qualitative agreement with phenomenological expectations~\cite{Beane:2014ora,Beane:2015yha,Parreno:2021ovq}.

It was pointed out early on~\cite{Iritani:2016jie} that the spectral representation for asymmetric correlators,
    $\left< \chi_A(t) \overline{\chi}_B(0) \right> = \sum_n Z_n^A [Z_n^B]^* e^{-E_n t}$,
can include excited-state contributions with opposite sign to the ground state when $A \neq B$.
In principle, these cancellations can lead to flat imaginary-time dependence at small $t$ that very slowly converges to a significantly different value at large $t$.
Evidence that this could be a practical concern grew dramatically when the Mainz group~\cite{Francis:2018qch,Green:2021qol}, the sLapHnn Collaboration~\cite{Horz:2020zvv}, and the NPLQCD Collaboration~\cite{Amarasinghe:2021lqa,Detmold:2024iwz} performed variational studies of two-baryon systems involving symmetric correlation functions with sources and sinks both of the form $N(\bm{p}) N(-\bm{p})$.
All three groups found that the energy spectra extracted by variational methods under ground-state dominance assumptions differed significantly from the spectrum obtained from asymmetric correlators under the same assumptions.
While asymmetric correlator results suggest $m_\pi \sim 800$ MeV leads to a bound dineutron and a deuteron somewhat more deeply bound than in nature, results from variational methods do not provide evidence for bound states in either channel.

\begin{wrapfigure}{r}{0.46\textwidth}
\vspace{-20pt}
\begin{center}
    \includegraphics[width=0.45\textwidth]{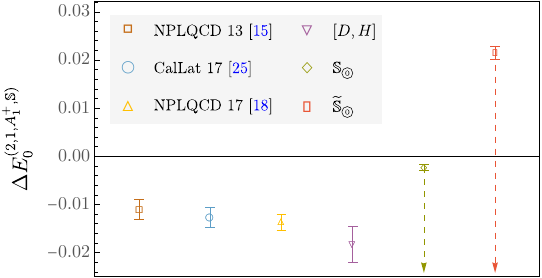}
\end{center}
\vspace{-20pt}
\caption{Comparison of asymmetric correlator estimates~\cite{NPLQCD:2012mex,Berkowitz:2015eaa,Wagman:2017tmp} with variational bounds~\cite{Amarasinghe:2021lqa} for dineutron FV energy shifts. All results used the same ensemble of gauge fields.  Reproduced from Ref.~\cite{Amarasinghe:2021lqa}.
\label{fig:variational}
}\vspace{-10pt}
\end{wrapfigure}
Variational methods employing ground-state energy estimators based on solving a generalized eigenvalue problem (GEVP)~\cite{Luscher:1990ck,Blossier:2009kd} provide rigorous upper bounds on the true ground-state 
energy, even for small imaginary times.
The $n$-th excited-state energy estimator provided by GEVP methods similarly provides an upper bound on the $n$-th excited-state energy of the true spectrum~\cite{Fleming:2023zml,Detmold:2024iwz}.
This means that variational results provide a qualitative improvement over asymmetric correlator results: rigorous one-sided bounds on the true spectrum.
However, these rigorous variational bounds do not exclude the earlier results from asymmetric correlators, see Fig.~\ref{fig:variational}.
The GEVP results directly provide a model of the spectrum in which the asymmetric 
correlator results arise via opposite-sign cancellations of ground- and excited-state contributions~\cite{Amarasinghe:2021lqa}; however, it is also possible to construct a model of the spectrum in which the effective mass from an asymmetric correlator converges to the true ground-state energy much faster than GEVP results obtained from a correlator matrix involving the same asymmetric correlator as an off-diagonal entry~\cite{Amarasinghe:2021lqa}.
Further explorations involving different interpolating operators~\cite{Detmold:2024iwz}, including a complete basis of local six-quark operators, have not significantly improved the lower bounds provided by those involving only $N(\bm{p}) N(-\bm{p})$ operators.

However, Hilbert space in LQCD is infinitely large~\cite{Kogut:1974ag} and cannot be searched exhaustively.
Although further explorations could provide increased confidence in some sense that ground-state saturation has been achieved in variational results,
is not clear how variational methods can be used to rigorously exclude the hypothesis that the earlier asymmetric correlator results are correct.

\section{Lanczos Algorithm for LQCD Spectroscopy}\label{sec:Lanczos}

The effective mass defined in Eq.~\eqref{eq:eff_mass} can be alternately viewed as the estimator associated with applying the power-iteration algorithm to find the largest eigenvalue of the transfer matrix~\cite{Wagman:2024rid}.
To see this, note that applying the power-iteration method~\cite{VonMises:1929} to find the largest eigenvalue of a Hilbert-space operator $T$ using an initial vector $\ket{\psi}$ corresponds to iteratively defining a ground-state eigenvector approximation $\ket{b_k} \propto T \ket{b_{k-1}}$ normalized such that $\braket{b_k}{b_k} = 1$ with the base case $\ket{b_1} = \ket{\psi}$.
The matrix element $\mbraket{b_k}{T}{b_k}$ then provides an iterative approximation that converges to the largest eigenvalue of $T$.
The transfer matrix, defined as the generator of Euclidean time translations by one lattice site, satisfies $C_\psi(t) = \mbraket{\psi}{T^{t/a}}{\psi}$.
Inserting a complete set of transfer-matrix eigenstates $\ket{n}$ gives
\begin{equation}
    C_\psi(t) = \sum_n \braket{\psi}{n} \mbraket{n}{T^{t/a}}{n} \braket{\psi}{n} = \sum_n |Z_n^\psi|^2 \lambda_n^{t/a},
\end{equation}
where $Z_n^\psi \equiv  \braket{\psi}{n}$. The spectrum of physical energies is defined by the transfer-matrix eigenvalues as $E_n = -(1/a) \ln \lambda_n$.
The ground-state energy estimator associated with the power-iteration method is therefore $-1/a$ times the log of the eigenvalue estimate $\mbraket{b_k}{T}{b_k}$,
\begin{equation}
    -\frac{1}{a} \ln \mbraket{b_k}{T}{b_k} = -\frac{1}{a} \ln  \left[ \frac{\mbraket{\psi}{T^{2k+1}}{\psi}}{\mbraket{\psi}{T^{2k}}{\psi}} \right] = -\frac{1}{a} \ln  \left[   \frac{C_\psi(t/a = 2k+1)}{C_\psi(t/a = 2k)} \right] = E(t/a = 2k),
\end{equation}
which establishes the equivalence of the effective mass and the power-iteration method.
The $O(e^{-\delta t})$ convergence of the effective mass corresponds to the standard $O(e^{-2k a\delta  }) = O([\lambda_1/\lambda_0]^{2k})$ convergence of the power-iteration method~\cite{Parlett}.

From this perspective, it is natural to ask whether more sophisticated algorithms from linear algebra could be used to determine transfer-matrix eigenvalues, and therefore the physical energy spectrum, more efficiently.
Methods such as the QR algorithm require an explicit representation of a matrix and cannot be straightforwardly applied to $T$ because it is an infinite-dimensional operator~\cite{Kogut:1974ag,Luscher:1976ms}.
However, the Lanczos algorithm~\cite{Lanczos:1950zz} is an appealing candidate because it involves iteratively constructing and explicitly diagonalizing low-rank approximations to a matrix.
The rank of the approximation is equal to the number of Lanczos iterations applied.
In computational linear algebra applications, only tens of iterations are often required to accurately approximate some eigenvalues of very large matrices, and convergence is observed to be fast for isolated eigenvalues, even when parts of the spectrum become dense~\cite{Parlett,Kuijlaars:2000,Saad:2011,Golub:2013,Garza-Vargas:2020}.
This suggests that Lanczos applied to an infinite-dimensional transfer matrix might provide rapidly convergent estimates for ground-state energies despite $T$ having an (infinitely) dense spectrum for energies of order $1/a$.

An algorithm for applying $m$ steps of Lanczos to an infinite-dimensional transfer-matrix $T$ with initial vector $\ket{\psi}$ that only requires $\{ C_\psi(0), C_\psi(a), \ldots, C_\psi((2m-1)a) \}$ as inputs is presented in Ref.~\cite{Wagman:2024rid}.
The starting point is that Lanczos eigenvalue approximations, called Ritz values, are obtained by diagonalizing a rank-$m$ matrix $T^{(m)}_{ij}$ with $i,j \in \{1,\ldots,m\}$ that for Hermitian $T$ takes the tridiagonal form $T^{(m)}_{ij} = \alpha_j \delta_{ij} + \beta_j [\delta_{i(j-1)} + \delta_{i(j+1)} ]$.
Constructing $T^{(m)}_{ij}$ only explicitly requires the $2m$ scalars $\{\alpha_1,\beta_1,\ldots,\alpha_m,\beta_m\}$ and not of any infinite-dimensional vectors or operators.
The usual three-term recurrence that defines Lanczos vectors, $T\ket{v_j} = \alpha_j\ket{v_j} + \beta_{j}\ket{v_{j-1}} + \beta_{j+1} \ket{v_{j+1}}$, can be used to define recursion relations for $\alpha_{j+1}$ and $\beta_{j+1}$ in terms of matrix elements computable at step $j$.
Defining $A_j^p \equiv \mbraket{v_j}{T^p}{v_j}$ and $B_j^p \equiv \mbraket{v_j}{T^p}{v_{j-1}}$, the required recursions are
\begin{equation} \label{eq:algo_diag}
\begin{split}
    A_{j+1}^p &= \frac{1}{\beta_{j+1}^2} \left[ 
    A_j^{p+2} 
    + \alpha_j^2 A_j^p   
    + \beta_j^2 A_{j-1}^p - 2\alpha_j A_j^{p+1}  
    + 2\alpha_j \beta_j B_j^p
    - 2\beta_j B_j^{p+1}   \right],
    \end{split}
\end{equation}
and
$B_{j+1}^p = \frac{1}{\beta_{j+1}} \left[ 
    A_j^{p+1} 
    - \alpha_j A_j^p
    - \beta_j B_j^p \right]$,
where $\alpha_j = A_j^1$, $\beta_{j+1} = \sqrt{A_j^2 - \alpha_j^2 - \beta_j^2}$, $\beta_1 = B_1^p = 0$, and $p \in \{1, \ldots 2(m-j)+1 \}$. 
Similar recursions are discussed in Ref.~\cite{DeMeo:1998}. 

This version of the Lanczos algorithm is inadequate for analyzing the noisy Monte Carlo estimators of correlators arising in LQCD. In particular, $\beta_{j+1} = \sqrt{A_j^2 - \alpha_j^2 - \beta_j^2}$ only results in $\beta_{j+1} \in \mathbb{R}$ if the data can be exactly described by a spectral representation with positive energies, which is violated by noise.
Allowing for $\beta_{j+1} \in \mathbb{C}$ instead provides a realization of the oblique Lanczos algorithm suitable for approximating eigenvalues of non-Hermitian matrices~\cite{Saad:1981}.
In this case, $T^{(m)}_{ij}$ can have complex eigenvalues. Defining energies from logarithms of eigenvalues, and even identifying the largest eigenvalue, can therefore have ambiguities when applied to noisy correlators discussed further below.
Different oblique Lanczos conventions can be employed that for example make $T^{(m)}_{ij}$ real with antisymmetric off-diagonal entries instead of complex and symmetric, but the Ritz values obtained as eigenvalues of $T^{(m)}_{ij}$ are convention-independent.

The Ritz values obtained by applying oblique Lanczos to noisy correlators turn out to be numerically identical to the polynomial roots obtained by applying Prony's method~\cite{Prony}---previously applied to LQCD correlators in Refs.~\cite{Fleming:2004hs,Lin:2007iq,Fleming:2009wb,Beane:2009kya,Fischer:2020bgv}---to the same data. I didn't know this at the time of Lattice 2024, where I presented a figure showing Lanczos results apparently converging faster than results from Prony's method. A few weeks later, George Fleming, Daniel Hackett, and I realized that there must be an error in my implementation of Prony's method due to apparent non-monotonicity. Upon fixing the error, we learned that Prony roots are numerically identical to the Ritz values from Lanczos. In August, v2 of Ref.~\cite{Wagman:2024rid} noted the coincidence between Lanczos and Prony results and referenced ongoing work to understand it more fully.
In November, the Bonn group posted Ref.~\cite{Ostmeyer:2024qgu}, which extends the observations made in v2 of Ref.~\cite{Wagman:2024rid} by providing an analytic proof of the correspondence between Ritz values and Prony roots. A few days later, the Tata group posted a proof of the same correspondence~\cite{Chakraborty:2024scw}.

The main utility of the Lanczos framework is that it provides useful formal technology in addition to a way to calculate the Ritz values (= Prony roots). In particular, it provides at least three novel features that are not apparent from the perspective of Prony's method:  1) theoretical bounds on convergence, 2) mathematically well-understood methods for identifying and removing spurious eigenvalues, and 3) directly calculable two-side error bounds.

Convergence of Lanczos is quantified by the Kaniel-Paige-Saad (KPS) bound~\cite{Kaniel:1966,Paige:1971,Saad:1980} on the difference between the true ground-state $T$ eigenvalue $\lambda_0$ and the Ritz value $\lambda_0^{(m)}$ after $m$ iterations,
\begin{equation}
    \frac{\lambda_0 - \lambda_0^{(m)}}{\lambda_0} \leq  \left[ \frac{\tan \arccos Z_0}{T_{m-1}(2 e^{a\delta} - 1)} \right]^2 \approx \frac{4 (1 - Z_0^2)}{Z_0^2} \times \begin{cases} e^{-2 (m-1) a\delta }    &  a\delta \gg 1 \\
    e^{-4(m-1)\sqrt{a\delta}} &  a\delta \ll 1 
    \end{cases}.
\end{equation}
For large gaps, Lanczos therefore has similar $e^{-2m a \delta}$ convergence as the power-iteration method. For small gaps, which arise generically near the continuum limit and especially for systems with relatively dense spectra where power-iteration convergence is slow, Lanczos achieves exponentially faster convergence proportional to $e^{-4 m \sqrt{a \delta}}$.
Analogous, although more complicated, bounds prove that $\lambda_1^{(m)}$ and other Ritz values provide estimators for excited-state energies that converge at similar rates.
However, the KPS bound does not apply to oblique Lanczos and is only applicable to LQCD correlators in the infinite-statistics limit. Further, since it involves the exact excitation gap and ground-state overlap the right-hand-side of the KPS bound is not directly calculable in practice.

In linear algebra applications to finite matrices, it has been shown that roundoff errors lead to the appearance of spurious eigenvalues that do not converge while not spoiling the convergence of non-spurious Ritz values~\cite{Paige:1971,Parlett,Cullum:1981,Cullum:1985,Kalkreuter:1995vg,Elsner:1999}.
The appearance of these spurious eigenvalues can be traced to numerical loss of orthogonality between Ritz vectors after multiple Lanczos iterations~\cite{Paige:1971,Parlett}.
Similar phenomena appear in applications to noisy correlator data~\cite{Wagman:2024rid}.
Although state-of-the-art methods for avoiding spurious eigenvalues using selective reorthogonalization~\cite{Parlett:1979} cannot be straightforwardly applied to LQCD correlator analysis, there is a method for identifying and filtering out spurious eigenvalues in a post-processing step developed by Cullum and Willoughby~\cite{Cullum:1981,Cullum:1985}.
Applying a bootstrap generalization of the Cullum-Willoughby (CW) test is sufficient to remove not only most complex eigenvalues (which can be identified as spurious simply by having significantly non-zero imaginary parts) but also real eigenvalues that would underestimate exact ground-state energies in solvable systems or predict violations of QCD inequalities like $M_N \geq m_\pi$~\cite{Weingarten:1983uj,Vafa:1983tf}.
The same spurious eigenvalues manifest as unphysical Prony roots in applications of Prony's method and have limited its practical application to LQCD correlators to $m \leq 4$~\cite{Lin:2007iq,Fleming:2009wb,Beane:2009kya,Fischer:2020bgv}.
After applying the CW test, it is feasible to obtain accurate energy estimates from oblique Lanczos even after $m \sim 50$ iterations.
Further, the SNR for Lanczos energy estimators does not inherit the same exponential SNR degradation as $C(t)$ or $E(t)$ if the CW test is applied and instead approaches an asymptotically constant value for large iteration counts.

\begin{figure}[t!]
                \centering
                \includegraphics[width=0.37\textwidth]{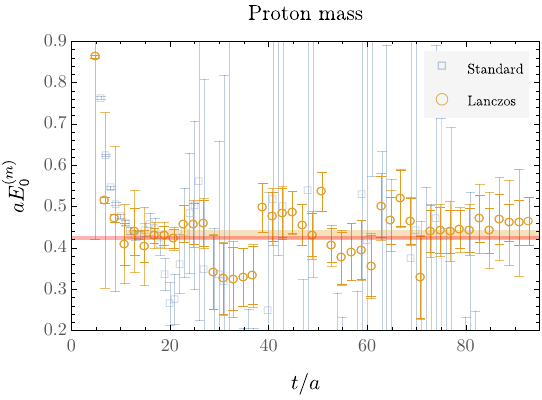}
                \includegraphics[width=0.48\textwidth]{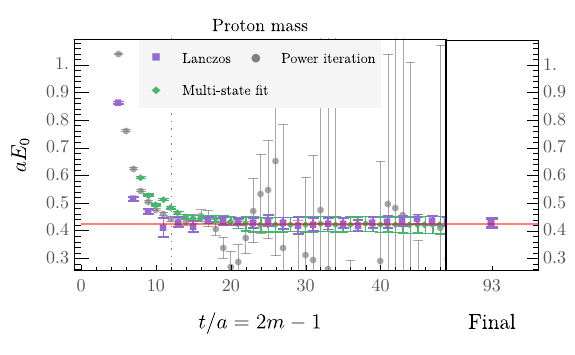}
            \caption{\label{fig:lanczos}
            Lanczos energy estimators for the nucleon mass using the basic sample-mean definition, left, and a bootstrap-median definition, right, are compared with the effective mass and multi-state fit results. Reproduced from Ref.~\cite{Wagman:2024rid} v1, left, and v3, right. Outer error bars, left, show residual-bound windows.
            }\vspace{-10pt}
\end{figure}
The appearance of asymptotically constant SNR can be associated with the appearance of correlations between energy estimators $E_k^{(m)}$ that grow with $m$ and suggest that Lanczos results are more analogous to multi-state fit results to the whole correlator $\{C(0), \ldots, C((2m-1)a)\}$ rather than a temporally localized estimator that would be particularly sensitive to $C((2m-1)a)$. These correlations are not manifestly visible using Lanczos energy estimators derived simply from sample-mean correlators. Employing outlier-robust estimators defined as medians over bootstrap resampled Ritz values provides a more precise and more accurate estimator because the outliers are primarily misidentified noise artifacts rather than part of the physical distribution~\cite{Wagman:2024rid,Hackett:2024nbe}. Using a bootstrap median definition, the iteration-by-iteration variance decreases significantly and correlations growing with $m$ can be clearly observed~\cite{Wagman:2024rid,Ostmeyer:2024qgu,Chakraborty:2024scw,Hackett:2024nbe}. This gives bootstrap-median Lanczos energy estimators the appealing feature that there is no need to perform a fit in order to extract all available information; simply picking a single iteration in the highly-correlated large $m$ region is sufficient to obtain precision that is similar to that achieved by either standard multi-state correlator fits or by fitting the full range of converged sample-mean Lanczos results, as shown for a LQCD nucleon correlator in Fig.~\ref{fig:lanczos}.
In particular, this circumvents standard worries about both covariance matrix estimation and sensitivity to a variety of fitting choices and hyperparameters.

Although the CW test implemented in Ref.~\cite{Wagman:2024rid} does introduce its own set of hyperparameters, their variation tends to be negligible unless it leads to obvious failures associated with classifying a physical eigenvalue as spurious, e.g.~disappearance of a Ritz value that is precisely determined with other hyperparameters.
Significant further progress in understanding spurious-eigenvalue filtering was made in the months after Lattice 2024 through the introduction of the ``ZCW test'' in Ref.~\cite{Hackett:2024nbe}. The ZCW shares the essential formal properties of the CW test, while providing a physical picture in which noise leads to the appearance of ``spurious states'' that are approximately orthogonal to the initial state but mix with Ritz vectors due to noise. 
Spurious states are defined as states with initial-state overlaps smaller than a threshold $\varepsilon_{\rm ZCW}$, set e.g.~by taking the minimum overlap  $Z_{\rm min}^2$ appearing in the first few iterations before spurious states emerge and setting $\varepsilon_{\rm ZCW} = Z_{\rm min}^2 / F_{\rm ZCW}$. The only hyperparameter $F_{\rm ZCW}$ enters in specifying how much lower the noise threshold should be than the minimum obviously physical overlap; $F_{\rm ZCW} \sim 10$ is suitable for many examples.

Perhaps the most important Lanczos feature is that there is a residual bound guaranteeing that at least one true eigenvalue exists within a two-sided window around each Ritz value whose size can be directly calculated. 
For symmetric Lanczos, this bound takes the form~\cite{Parlett}
\begin{equation}
  \min_{\lambda \in \{\lambda_n\}} |\lambda_k^{(m)} - \lambda |^2 \leq B_k^{(m)} \equiv |\beta_{m+1}|^2 \, |\omega_{mk}^{(m)}|^2,
\end{equation}
where $\omega_k^{(m)}$ is the $k$-th eigenvector of $T^{(m)}_{ij}$.
The oblique case for noisy correlators is only modified by the appearance of an additional calculable ratio of Ritz- and Lanczos-vector norms on the right-hand-side~\cite{Wagman:2024rid}.
Note that this bound does not guarantee that $\lambda_0^{(m)}$ has converged to within $\sqrt{B_0^{(m)}}$ of the true ground-state $\lambda_0$; however, it does guarantee that some true eigenvalue $\lambda_n$ is within $\left[ \lambda_0^{(m)} - \sqrt{B_0^{(m)}}, \lambda_0^{(m)} + \sqrt{B_0^{(m)}}\right]$.
This provides a qualitative advance over variational methods that can only rigorously provide one-sided bounds.
Further, residual bounds are computable for both symmetric and asymmetric correlators.
This could be critically usefully in particular for testing the validity of ground-state saturation assumptions in the analyses of nuclear correlation functions discussed in Sec.~\ref{sec:nuclei}.
More generally, residual bounds provide rigorous information about LQCD spectra without requiring any assumptions about how well ground-state convergence has been achieved at finite imaginary times or how well interpolating operators overlap with particular states.
This provides a new (two-sided) window on QCD whose exploration is just beginning.

\begin{acknowledgments}
I would like to thank the Organizers of Lattice 2024 and the Kenneth G. Wilson Award committee for the honor of giving this presentation and a very enjoyable conference.
I am deeply grateful to all of the collaborators, mentors, and friends that I have been fortunate to know in my time at the University of Washington, MIT, and Fermilab, and I especially thank my PhD advisor Martin~Savage. 
The work that I presented here was done in collaboration with Artur Avkhadiev, Phiala Shanahan, and Yong~Zhao on the Collins-Soper kernel, the NPLQCD Collaboration on multi-nucleon systems in LQCD, and Dan~Hackett on the Lanczos algorithm.
This manuscript has been authored by FermiForward Discovery Group, LLC under Contract No. 89243024CSC000002 with the U.S. Department of Energy, Office of Science, Office of High Energy Physics.
\end{acknowledgments}

\bibliographystyle{JHEP}
\bibliography{refs}

\end{document}